\documentclass[twocolumn,superscriptaddress,showpacs,preprintnumbers,amsmath,amssymb,prl]{revtex4}
\usepackage{graphicx}
\usepackage{dcolumn}
\usepackage{bm}

\begin{document}
\title{The nature of Ho magnetism in multiferroic HoMnO$_3$}

\author{S. Nandi}
\author{A. Kreyssig}
\author{L. Tan}
\author{J. W. Kim}
\author{J. Q. Yan}
\affiliation{Ames Laboratory, USDOE and Department of Physics and
Astronomy, Iowa State University, Ames, Iowa 50011, USA }

\author{J.~C.~Lang}
\author{D.~Haskel}
\affiliation{Advanced Photon Source, Argonne National Laboratory,
Argonne, Illinois 60439, USA }

\author{R.~J.~McQueeney}
\author{A.~I.~Goldman}
\affiliation{Ames Laboratory, USDOE and Department of Physics and
Astronomy, Iowa State University, Ames, Iowa 50011, USA }


\begin{abstract}
Using x-ray resonant magnetic scattering and x-ray magnetic circular
dichroism, techniques that are element specific, we have elucidated
the role of Ho\textsuperscript{3+} in multiferroic HoMnO$_{3}$. In
zero field, Ho\textsuperscript{3+} orders antiferromagnetically with
moments aligned along the hexagonal \textbf{c} direction below 40~K,
and undergoes a transition to another magnetic structure below
4.5~K. In applied electric fields of up to $1\times10^{7}$~V/m, the
magnetic structure of Ho\textsuperscript{3+} remains unchanged.
\end{abstract}

\pacs{75.25.+z, 75.47.Lx, 75.50.Ee, 75.80.+q, 77.80.-e} 

\maketitle

Magnetoelectric multiferroic compounds, systems which exhibit both
ferroelectric and magnetic effects within the same phase, have
attracted considerable attention due to the potential for
controlling electric polarization by an applied magnetic
field\cite{kimura_magnetic_2003} or, conversely, magnetic order
through an applied electric field\cite{lottermoser_magnetic_2004}.
Recently, such a mechanism has been proposed for hexagonal
HoMnO$_3$\cite{lottermoser_magnetic_2004}. Despite numerous studies,
however, the exact role that the Ho\textsuperscript{3+} ions play in
the magnetic response, and the details of the magnetic ordering of
the Ho\textsuperscript{3+} sublattices remain
unclear\cite{sugie_magnetic_2002, fiebig_JAP,
fiebig_determination_2000, lonkai_magnetic_2002,
munoz_evolution_2001, brown_neutron_2006}.

Below \textit{T}$_\mathrm{C}$ = 875~K, the ferroelectric phase of
HoMnO$_3$ possesses \textit{P6}$_3$\textit{cm} symmetry and a
polarization \textit{P}$_\mathrm{C}$ = 5.6${\mu}$C/cm$^2$ along the
hexagonal \textbf{c} axis\cite{coeure_1996}. The
Mn\textsuperscript{3+} moments order antiferromagnetically within
the \textbf{a-b} plane below the N\'eel temperature,
\textit{T}$_\mathrm{N}$ = 76~K, and undergo reorientation
transitions at approximately 40~K and
5~K\cite{fiebig_determination_2000, vajk_magnetic_2005}. There have
been several investigations of the role of Ho\textsuperscript{3+} in
the magnetic ordering of this compound, but with contradictory
results. For clarity in the discussion that follows, we list the
various magnetic symmetries and their features in Table
\ref{tab:groups} and refer to the associated magnetic configurations
for Ho\textsuperscript{3+} by their magnetic representation.

\begin{table*}
\centering \caption{Magnetic representations for HoMnO$_3$ 
according to Ref.~2 and Ref.~7 for moments along the \textbf{c} 
direction. Additional magnetic representations, $\Gamma_5$ and 
$\Gamma_6$, allow moments only in the \textbf{a-b} plane and only 
$\Gamma_6$ yields magnetic intensity for $(0~0~l)$ and $(h~0~l)$ 
reflections with \textit{l}~odd.  The atomic positions for Ho are 
given in brackets. $[z, +, -]$ depict $z_{2a} = 0.273$, 
$+\mu^c_{2a}$, $-\mu^c_{2a}$, and $z_{4b} = 0.231$, 
$+\mu^c_{4b}$, $-\mu^c_{4b}$ for the Wyckoff sites 
Ho~(2\textit{a}) and Ho~(4\textit{b}), 
respectively\cite{munoz_evolution_2001}. The symbol $[0]$ labels 
no ordered magnetic moment at this site.} \label{tab:groups}
\begin{ruledtabular}
\begin{tabular}{c|cc|cccc|cc}
 & \multicolumn{2}{c}{Ho (2\textit{a})} \vline& \multicolumn{4}{c}{Ho (4\textit{b})} \vline& \multicolumn{2}{c}{Magnetic intensity}\\
\cline{2-9}$\begin{matrix}\mathrm{Magnetic} \\
\mathrm{representation}\end{matrix}$ &
$\begin{pmatrix}0\\0\\z\end{pmatrix}$ &
$\begin{pmatrix}0\\0\\z+1/2\end{pmatrix}$ &
$\begin{pmatrix}1/3\\2/3\\z\end{pmatrix}$ &
$\begin{pmatrix}2/3\\1/3\\z\end{pmatrix}$ &
$\begin{pmatrix}1/3\\2/3\\z+1/2\end{pmatrix}$  &
$\begin{pmatrix}2/3\\1/3\\z+1/2\end{pmatrix}$ &
$\begin{matrix}(0~0~l),\\l~odd\end{matrix}$ &
$\begin{matrix}(h~0~l),\\l~odd\end{matrix}$\\
 \hline
$\Gamma_1$&0&0&$+$&$-$&$-$&$+$&no&yes\\
$\Gamma_2$&$+$&$+$&$+$&$+$&$+$&$+$&no&no\\
$\Gamma_3$&$+$&$-$&$+$&$+$&$-$&$-$&yes&yes\\
$\Gamma_4$&0&0&$+$&$-$&$+$&$-$&no&no\\
\end{tabular}
\end{ruledtabular}
\end{table*}

Mu\~noz \emph{et al.}\cite{munoz_evolution_2001} proposed that the
magnetic representation is $\Gamma_1$ in the temperature range
24.6~K to 1.7~K based upon neutron powder diffraction measurements.
Lonkai \emph{et al.}\cite{lonkai_magnetic_2002} and Fiebig \emph{et
al.}\cite{fiebig_JAP, fiebig_determination_2000} proposed that from
40~K to 5~K the magnetic representation is $\Gamma_3$ and
transforms, below 5~K, to $\Gamma_1$ based on neutron diffraction
experiments and optical second harmonic generation (SHG)
experiments, respectively. Most recently, Brown \emph{et
al.}\cite{brown_neutron_2006} claimed that the magnetic
representation is $\Gamma_3$ from 40~K down to 2~K based on neutron
diffraction on single crystals and powders. All of the above studies
concluded that the Ho\textsuperscript{3+} moments order along the
hexagonal \textbf{c} axis. In their analysis of magnetoelectric
measurements, however, Sugie \emph{et al.}\cite{sugie_magnetic_2002}
proposed that the Ho\textsuperscript{3+} moments order
non-collinearly in the hexagonal \textbf{a-b} plane. Perhaps of
strongest interest is the proposal by Lottermoser \emph{et
al.}\cite{lottermoser_magnetic_2004} that the application of an
electric field changes the antiferromagnetic order of
Ho\textsuperscript{3+} to ferromagnetic order, with the
representation $\Gamma_2$, over the temperature range from 2~K to
76~K, based on SHG and optical Faraday rotation experiments.

To resolve the contradictions regarding the magnetic order of
Ho\textsuperscript{3+} below 40~K in zero field, and to investigate
the nature of the Ho\textsuperscript{3+} magnetic ordering in an
applied electric field, we have performed x-ray resonant magnetic
scattering (XRMS) and x-ray magnetic circular dichroism (XMCD)
studies of HoMnO$_3$ at the Ho L$_\mathrm{III}$ absorption edge
(\textit{E} = 8.071 keV). In the XRMS experiment, the scattering
process involves an intermediate state which arises from either
dipole (E1) allowed (2p {--} 5d) or quadrupole (E2) allowed (2p {--}
4f) electronic excitations\cite{gibbs_polarization_1988,
hannon_x-ray_1988}. In ordered rare-earth magnetic materials, the
technique is sensitive to the magnetization density through either
the 4f electronic states directly (E2) or indirectly through the 4f
{--} 5d exchange interaction (E1). Of importance here is that since
this technique is element specific, we can probe the magnetic
structure associated with the Ho\textsuperscript{3+} moments
directly.  In the closely related XMCD measurements, the signal is
defined as the difference in the absorption of circularly polarized
x-rays with the helicity parallel and antiparallel to the sample
magnetization\cite{Jonathan}. Since XMCD measurements are also
performed at the absorption edges of elements of interest they can
be viewed as measurements of the net magnetization for a specific
elemental constituent of a magnetic compound, for example, measuring
the contribution of Ho\textsuperscript{3+} to any ferromagnetic
response of the sample.

Single crystals of HoMnO$_3$ were grown using a floating zone 
method and prepared with a polished surface perpendicular to the 
crystallographic \textbf{c} axis. The magnetic properties (e.g. 
details of the complex magnetic phase diagram) agree well with 
reported results\cite{vajk_magnetic_2005, 
lorenz_field-induced_2005}. The XRMS experiment was performed on 
the 6ID-B beamline at the Advanced Photon Source at the Ho 
L$_\mathrm{III}$ absorption edge. The sample was mounted on the 
cold-finger of a displex cryogenic refrigerator with the 
\textbf{a\textsuperscript{*}-c\textsuperscript{*}} reciprocal 
plane coincident with the scattering plane.  With the incident 
beam polarized perpendicular to the scattering plane 
($\sigma$-polarized), measurements of the magnetic scattering 
were performed at the E1 and E2 resonances in the rotated 
($\sigma$-$\pi$) scattering channel\cite{cryst_x-ray_1996}. 
Pyrolytic graphite (0~0~6) was used as a polarization analyzer to 
suppress the charge background (unrotated) relative to the 
magnetic scattering signal. The XMCD measurements were performed 
on the 4ID-D beamline at the Advanced Photon Source by modulating 
the x-ray helicity at 11.3 Hz with a thin diamond phase retarder 
and using a lock-in amplifier to detect the related modulation in 
the absorption coefficient. Spectra were recorded from a 
20~${\mu}$m thin single crystal coated with carbon as electrodes 
and on a powder sample spread over several layers of tape and 
then sandwiched between two layers of conducting aluminized mylar 
(total sandwich thickness of~280~${\mu}$m). The samples were 
mounted on the cold finger of a horizontal field cryomagnet.

\begin{figure}
\begin{center}
\includegraphics[clip, width=0.45\textwidth]{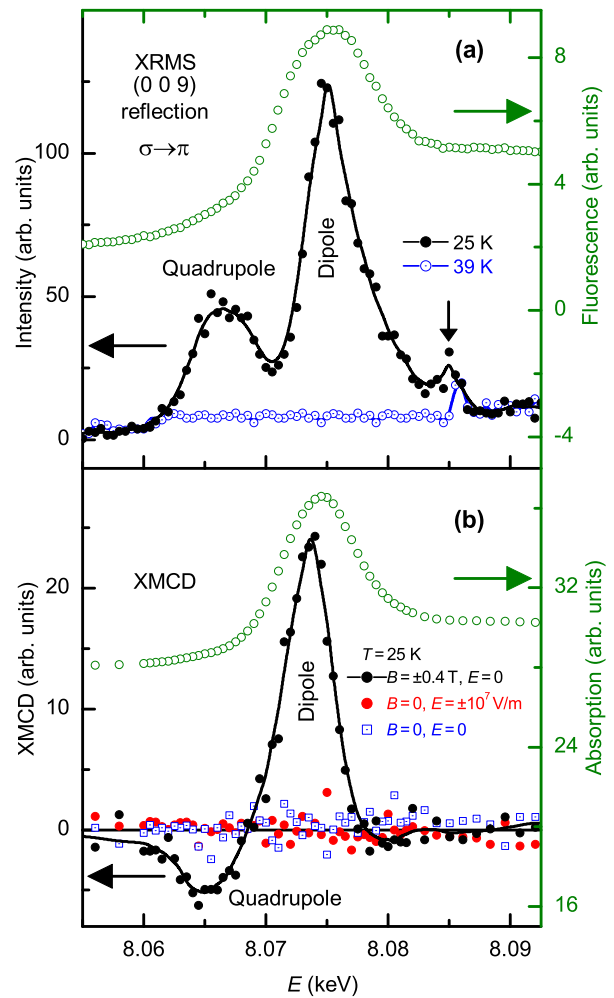}
\caption{(a) Energy scans for the magnetic (0~0~9) reflection and
the fluorescence spectra through the Ho
  $L_{\mathrm{I\!I\!I}}$ absorption edge. The solid
lines are guides to the eye. The small peak marked by a vertical
arrow is due to multiple charge scattering. (b) The XMCD and
absorption spectra at the Ho $L_{\mathrm{I\!I\!I}}$ edge with
applied magnetic, electric fields and without any fields, measured
on a powder sample at 4ID-D. The solid line is a guide to the eye.}
\label{fig:Resonance_enhancement}
\end{center}
\end{figure}

We first turn our attention to the magnetic structure of
Ho\textsuperscript{3+} in the absence of an applied electric field.
Fig.~\ref{fig:Resonance_enhancement}(a) shows energy scans through
the Ho L$_\mathrm{III}$ absorption edge at the (0~0~9) reciprocal
lattice point which is forbidden for charge scattering. Below 39~K,
we observed two resonance peaks which correspond to the quadrupole
resonance (just below the Ho L$_\mathrm{III}$ absorption edge) and
the dipole resonance (just above the Ho L$_\mathrm{III}$ absorption
edge). Recent XRMS work on TbMnO$_3$\cite{mannix_07}, has shown that
resonant scattering features can involve both charge and magnetic
contributions in the more complex perovskite multiferroic compounds.
In hexagonal HoMnO$_{3}$, the predominant magnetic origin of the
resonant scattering is supported by the following points.  First, we
note the close proximity in energy between the XRMS resonances
(Fig.~\ref{fig:Resonance_enhancement}(a)) and the quadrupole and
dipole features in the XMCD spectrum of the powder sample in an
applied magnetic field of 0.4 Tesla
(Fig.~\ref{fig:Resonance_enhancement}(b)). The magnetic origin of
the XMCD features was confirmed by the observation that the spectrum
flips upon reversal of the applied magnetic field. In addition, fits
to the XRMS integrated intensities for (0~0~\emph{l}) (\emph{l} odd)
diffraction peaks measured at each resonance are consistent with the
expected angular and polarization dependence of the dipole and
quadrupole resonant scattering cross-sections,
respectively\cite{cryst_x-ray_1996,nandi2008,detlefs_1997}.

\begin{figure}
\begin{center}
\includegraphics[clip, width=0.43\textwidth]{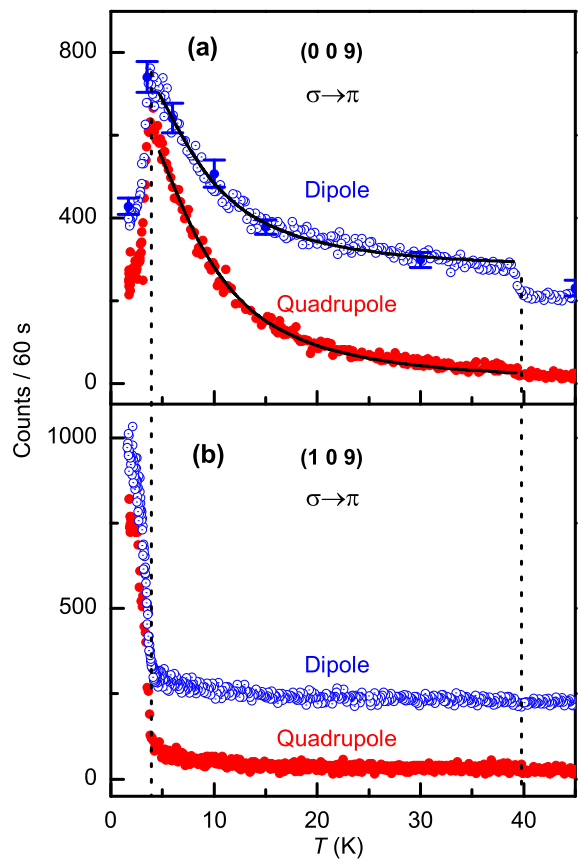}
\caption{Temperature dependence of the magnetic peak intensity for
the (0~0~9) and (1~0~9) reflections, measured for both the dipole
and quadrupole resonances. The dipole resonance data are vertically
displaced for clarity. Solid blue circles represent the integrated
intensity of peaks measured at selected temperatures. The solid
black lines are fits to the data in the intermediate magnetic phase
as described in the text.} \label{fig:Tdep1}
\end{center}
\end{figure}

Fig.~\ref{fig:Tdep1} shows the measured peak intensities for two
characteristic magnetic reflections, (0~0~9) and (1 0 9), as a
function of temperature from 1.7~K to 45~K. Also shown are the
integrated intensities measured in ${\theta}$-scans (rocking scans)
at several temperatures for the (0~0~9) reflection. Due to strong
local heating effects from the incident beam, significant
attenuation of the undulator beam (${\geq}$~98~\%) was required to
accurately measure the temperature dependence over the entire
temperature range. Above \textit{T}~=~39~K, we find no signal at
these magnetic peak positions. Just below 39~K there is an abrupt
increase in both the dipole and quadrupole diffraction peak
intensities, followed by a smooth increase with decreasing
temperature down to 4.5~K. The temperature of the onset of magnetic
Ho\textsuperscript{3+} ordering agrees well with the kink in
magnetization data and the Mn\textsuperscript{3+} spin
re-orientation transition as found by Vajk \emph{et
al.}\cite{vajk_magnetic_2005}. At 4.5~K, changes in the intensities
of the (0~0~9) and (1~0~9) magnetic peaks signal a change in the
magnetic ordering of Ho\textsuperscript{3+} in this low temperature
phase (LTP), from the intermediate temperature phase (ITP) for 4.5~K
{\textless} T {\textless} 39~K.

To determine the moment direction in the intermediate phase, we
measured the intensity of the off-specular reflections, (3 0 9) 
and ($\overline{3}$ 0 9), at \textit{T}~=~12~K. While the 
magnetic structure factor is the same for both reflections, the 
angular dependence of the dipole resonant scattering cross 
section\cite{cryst_x-ray_1996} is different providing strong 
sensitivity to the moment direction. The experimentally observed 
intensity ratio $I (3~0~9)/ I (\overline{3}~0~9)$ is $76\pm14$. 
For the Ho\textsuperscript{3+} moment oriented along the 
hexagonal \textbf{c} axis, the calculated intensity ratio 
\textit{I} (3 0 9)/\textit{I}($\overline{3}$~0~9) = 86.3, while 
for moments lying in the \textbf{a-b} 
plane,~\textit{I}~(3~0~9)/\textit{I} ($\overline{3}$~0~9)~=~0.32. 
Thus, within experimental error, the Ho\textsuperscript{3+} 
moments lie along the hexagonal \textbf{c} axis.

As identified in previous work, the magnetic unit cell is the same
as the chemical unit cell\cite{munoz_evolution_2001}. In order to
determine the appropriate magnetic
representation\cite{bertaut_representation_1968} we must look into
details of the six magnetic representations that are possible for
the crystallographic space group \textit{P6$_3$cm}, listed in
Ref.~7. Only four magnetic representations (from $\Gamma_1$ to
$\Gamma_4$) are compatible with Ho\textsuperscript{3+} magnetic
moments along the hexagonal \textbf{c} direction and are described
in Table~\ref{tab:groups}. From the last columns of
Table~\ref{tab:groups}, we see that only the magnetic representation
$\Gamma_3$ yields non-zero intensity for (0 0 \textit{l})
reflections with \textit{l} odd.  While details will be presented
elsewhere\cite{nandi2008}, here we note that at 6~K, the measured
integrated intensities of a series of (0~0~\textit{l}) reflections
are consistent with calculated values for both dipole and quadrupole
resonant scattering based on the magnetic representation $\Gamma_3$
and yields an ordered magnetic moment for the Ho (2\textit{a}) site
twice that of the Ho~(4\textit{b}) site at 6~K.

The temperature dependence of the Ho\textsuperscript{3+} XRMS signal
in the ITP (as shown in Fig.~\ref{fig:Tdep1}(a)) exhibits a concave
curvature quite similar to what was observed in XRMS measurements of
orthorhombic multiferroic DyMnO$_3$, and ascribed to the induced
magnetic ordering of Dy\textsuperscript{3+}\cite{prokhnenko_07}. For
hexagonal HoMnO$_3$, this behavior may be explained with reference
to other systems with ground state quasidoublet crystal field
levels, split by an exchange field\cite{popova_high-resolution_2005,
sachidanandam_single-ion_1997, zheludev_x-ray_1996}. The non-Kramers
Ho\textsuperscript{3+} ions in HoMnO$_3$ are at positions of
trigonal symmetry which typically leads to singlets with very small
zero-field splitting, forming such a
quasidoublet\cite{PhysRev.188.539, hardy_magnetism_2006}. As a first
attempt, if we assume an effective two level system with a splitting
${\Delta}_\mathrm{{eff}}$, the temperature dependence in the ITP,
should goes as $(\mathrm{tanh}\frac{\Delta_{\mathrm{eff}}}{2kT})^2$
\cite{popova_high-resolution_2005, sachidanandam_single-ion_1997}.
Fits to the temperature dependence of the dipole and quadrupole
resonant scattering in the ITP yields
${\Delta}_\mathrm{{eff}}=(1.3~\pm~0.2)$~meV, consistent with the
low-energy crystal electric field transition of
Ho\textsuperscript{3+} of 1.5~meV observed in neutron scattering
measurements below 40K\cite{vajk_magnetic_2005, zhou_specific_2006}.
The ordering at the Ho\textsuperscript{3+} sites in the ITP may be
induced by changes in the anisotropic superexchange interactions
resulting from the reorientation of the Mn\textsuperscript{3+}
moments\cite{fiebig_02}.

\begin{figure}
\begin{center}
\includegraphics[clip, width=0.45\textwidth]{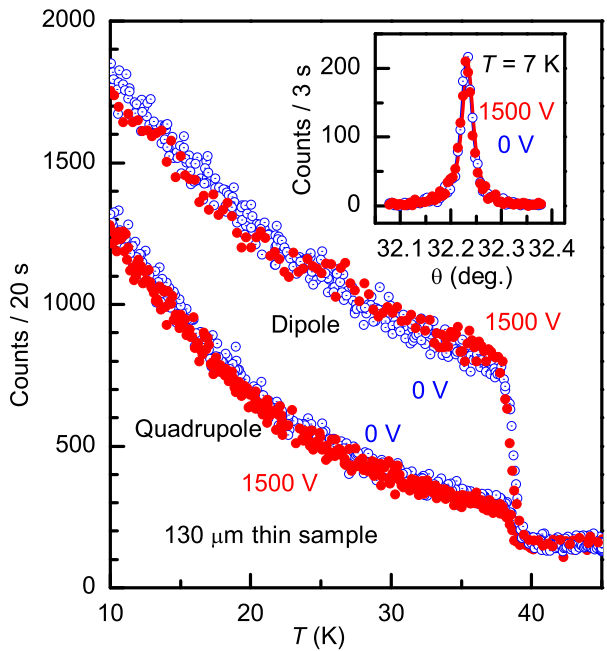}
\caption{Effect of an applied electric field on the (0 0 9) 
antiferromagnetic peak. Inset: rocking scans at 0~V and 1500~V for 
the quadrupole resonance at 7~K.} \label{fig:E-field_dep}
\end{center}
\end{figure}

We now turn to our investigation of the magnetic structure of the
low temperature phase (LTP). From Fig.~\ref{fig:Tdep1} we see that
the intensity of (0~0~9) decreases while the intensity of (1~0~9)
increases strongly below 4.5~K. Similar to our approach for the ITP,
we confirmed that the moments lie primarily along the \textbf{c}
axis by measuring the integrated intensities of the (2~0~9) and
($\overline{2}$~0~9) reflections at 2 K. The decrease in intensity
for the (0~0~9) reflection and increase in the (1~0~9) reflection
signal a transition from $\Gamma_3$ to $\Gamma_1$ (see Table
\ref{tab:groups}). An interesting consequence of this transition is
that the Ho (2\textit{a}) site can not order magnetically according
to the representation $\Gamma_1$. The finite intensity of the
(0~0~9) reflection below 4.5~K is likely due to residual beam
heating effects although we can not exclude a mixed
$\Gamma_3$/$\Gamma_1$ phase in this region.

After determining the magnetic structure in zero field, we 
measured the temperature dependence of both dipole and quadrupole 
resonances in an applied electric field. For these measurements, 
a thinned (130~${\mu}$m) sample coated with silver (as 
electrodes) was used. Fields of up to $1\times10^{7}$~V/m, well 
above the saturation value reported by Lottermoser \emph{et 
al.}\cite{lottermoser_magnetic_2004}, were obtained for an applied 
voltage of 1500~V. Fig.~\ref{fig:E-field_dep} shows that there is 
no difference between the temperature dependence of the peak 
intensity in zero and the maximum applied electric field. 
Further, the inset to the Fig.~\ref{fig:E-field_dep} shows that 
there are no gross structural changes in this applied field since 
the peak position and the full width at half maximum remain same. 
We conclude that there is no change of the antiferromagnetic 
structure of Ho\textsuperscript{3+} in an applied electric field 
in the ITP.

Since the XRMS measurements probe only antiferromagnetic order and
are not sensitive to small ferromagnetic components of the ordered
magnetic moment, we have performed XMCD measurements, at the Ho
L$_\mathrm{III}$ edge, on single crystal and powder samples. As
illustrated for the powder sample in
Fig.~\ref{fig:Resonance_enhancement}~(b), the XMCD signal is 
unaffected by applied electric fields of up to approximately 
$1\times10^{7}$~V/m for either sample, for temperatures between 
8~K and 80~K. We conclude that Ho\textsuperscript{3+} is not 
responsible for the reported ferromagnetic response of 
HoMnO$_{3}$ in an applied electric field.

\begin{acknowledgments}
We are indebted to D. S. Robinson, A. Kracher, and A. Barcza for
help during experiments. We thank J. Schmalian and B. N. Harmon for
helpful discussions. The work at Ames Laboratory and at the MU-CAT
sector was supported by the US DOE under Contract No.
DE-AC02-07CH11358. Use of the Advanced Photon Source was supported
by US DOE under Contract No. DE-AC02-06CH11357.
\end{acknowledgments}

\bibliographystyle{apsrev}
\bibliography{homno3_4}

\end{document}